\documentclass[letterpaper,12pt]{JHEP3}
\usepackage{amsmath}
\usepackage{amssymb}
\usepackage{graphicx}
\usepackage{subfig}
\usepackage{bbm}
\usepackage{epsfig}
\usepackage{yfonts}
\def\({\left(} \def\){\right)}
\def\[{\left[} \def\]{\right]}

\newcommand{\be}{\begin{equation}}
\newcommand{\ee}{\end{equation}}
\newcommand{\bea}{\begin{eqnarray}}
\newcommand{\eea}{\end{eqnarray}}
\newcommand{\ba}{\begin{eqnarray}}
\newcommand{\ea}{\end{eqnarray}}

\newcommand{\beq}{\begin{equation}}
\newcommand{\eeq}{\end{equation}}
\newcommand{\beqa}{\begin{eqnarray}}
\newcommand{\eeqa}{\end{eqnarray}}
\newcommand{\beqar}{\begin{eqnarray*}}
\newcommand{\eeqar}{\end{eqnarray*}}

\preprint{arXiv:1111.0273 [hep-ph]}

\title{The Minimal Dimensionless Standard Model (MDSM) and its
  Cosmology}

\author{Latham Boyle, Shane Farnsworth, Joseph Fitzgerald
and Maitagorri Schade\\
\it
Perimeter Institute for Theoretical Physics, Waterloo, Ontario N2L 2Y5, Canada}

\abstract{Consider the minimal renormalizable extension of the
  Standard Model with purely dimensionless couplings, successful
  electroweak symmetry breaking (via the Coleman-Weinberg mechanism)
  and a see-saw mechanism for neutrino mass: we will call this the
  Minimal Dimensionless Standard Model (MDSM).  In fact, 3 closely
  related models fit the bill: MDSM$_{1}$, MDSM$_{2}$ and MDSM$_{3}$.
  We analyze the theoretical and observational constraints on these
  models.  We argue that, when they are minimally coupled to gravity,
  they can accomplish several important cosmological tasks (inflation,
  dark matter, leptogenesis) in a way that is economical, predictive
  and tightly woven into the fabric of known physics. One of the
  models (MDSM$_{3}$), which includes an extra $U(1)_{B-L}$ gauge
  symmetry, seems particularly promising.}

\begin{document}

%%%%%%%%%%%%%%%%%%%%%%%%%%%%%%%%%%%%%%%%%%%%%%%%%
\section{Introduction}
\label{intro}

Over the years, many physicists have been intrigued (for a variety of
different reasons) by the idea that the laws of physics might, at
fundamental level, be based on massless particles and dimensionless
couplings; and that masses and other dimensionful quantities might, in
some sense, be secondary or emergent.  This has led some researchers
to pay special attention to gauge field theories with purely
dimensionless coupling constants ({\it e.g.}\ \cite{Coleman:1973jx,
  Gildener:1976ih, Witten:1980ez}).  Although such theories possess
conformal symmetry at the classical level, this symmetry is generally
violated at the quantum level.  Nevertheless, these theories are
strictly renormalizable in the sense that ({\it e.g.}\ if we use
dimensional regularization) there are no ultraviolet divergences that
require counterterms with dimensionful coupling constants; the
dimensionless couplings just run logarithmically with scale.  Some
physicists have argued that the resolution of the standard model
``hierarchy problem'' may lie in the standard model's proximity to a
model with purely dimensionless couplings ({\it e.g.}\
\cite{Gildener:1976ih, Bardeen:1995kv}).

In the standard model of particle physics (which we here take to
include 3 right-handed neutrinos $\nu_{R}$), most of the coupling
constants {\it are} dimensionless.  Dimensionful couplings only appear
in two places in the Lagrangian: (i) in the Majorana mass term
$\nu_{R}^{T}\gamma^{0}\gamma^{2} M_{m}\nu_{R}$ for the right-handed
neutrinos; and (ii) in the quadratic self-coupling $m^{2}h^{\dagger}h$
of the Higgs doublet $h$.  (Here the $3\times3$ matrix $M_{m}$ and the
constant $m$ both have dimensions of mass.)  Is it possible that these
two dimensionful terms actually vanish, so that the Lagrangian only
contains dimensionless couplings?  The answer is no: this model (which
we shall call the ``Minimal Dimensionless Standard Model, Version 0''
or ``MDSM$_{0}$'') is ruled out.  These two dimensionful terms each
play an important phenomenological role in the standard model: the
$\nu_{R}^{T}\gamma^{0}\gamma^{2}M_{m}\nu_{R}$ term is responsible for
the ``see-saw mechanism,'' the best available explanation for the
smallness of the observed neutrino masses; and, crucially, the
$m^{2}h^{\dagger}h$ term is needed to generate spontaneous breaking of
electroweak symmetry.  (One might wonder whether it is possible to set
$m=0$, and still have electroweak symmetry breaking via the
Coleman-Weinberg mechanism \cite{Coleman:1973jx, Gildener:1976ih};
but, as we shall review in Sec.~\ref{BosonMassBound}, this doesn't
work in MDSM$_{0}$ \cite{Sher:1988mj}.)

Instead, let us ask for the minimal renormalizable {\it extension} of
the standard model that only contains dimensionless couplings in its
Lagrangian, but nevertheless exhibits a see-saw mechanism for neutrino
masses, and spontaneous electroweak symmetry breaking via the
Coleman-Weinberg mechanism.  In fact, there are three closely related
variants that fit the bill: we will call them MDSM$_{1}$, MDSM$_{2}$
and MDSM$_{3}$.  All three models have previously been discussed in
the literature: MDSM$_{1}$ in \cite{Meissner:2006zh}, MDSM$_{2}$ in
\cite{Meissner:2008gj, Latosinski:2010qm, AlexanderNunneley:2010nw},
and MDSM$_{3}$ in \cite{Iso:2009ss}.  (For other previous work on
dimensionless variants of the standard model, see
\cite{Hempfling:1996ht, Foot:2007as, Foot:2007ay, Foot:2007iy,
  Foot:2010av, Foot:2010et, Holthausen:2009uc}.)  In this paper, we
re-assess these three models and their cosmological consequences,
contributing a range of new results.

The layout of the paper is the following.  Sections \ref{MDSM} and
\ref{tree_level_masses} are intended to establish our notations and
conventions: in Section \ref{MDSM}, we specify the three models
MDSM$_{1}$, MDSM$_{2}$ and MDSM$_{3}$; and in Section
\ref{tree_level_masses} we present the tree-level mass spectra in
these theories, which we will need for our subsequent analysis.  In
Section \ref{ColemanWeinberg} we discuss spontaneous symmetry breaking
in these theories via the Coleman-Weinberg mechanism.  First, in
Subsection \ref{GildenerWeinberg}, we present a convenient formalism,
due to E.~Gildener and S.~Weinberg \cite{Gildener:1976ih}, for
extracting some reliable results about Coleman-Weinberg symmetry
breaking, without having to calculate the full
renormalization-group-improved effective potential.  Then in
Subsections \ref{BosonMassBound} and \ref{NeutrinoMassBound} we apply
this formalism to our three models to obtain lower bounds on the new
bosonic masses, and upper bounds on the heavy neutrino masses.

In Section \ref{inflation} we discuss inflation.  In our three models,
the inflaton candidate is the scalar field $\psi_{\parallel}$ which
rolls along the ``Gildener-Weinberg'' direction connecting the origin
in field space (the point of unbroken symmetry) to the
symmetry-breaking VEV.  In Subsection \ref{inflation_mass_relation},
we explain that inflation predicts a specific mass relation between
the heaviest bosonic and fermionic particles in these theories; in
Subsection \ref{perturbations}, we present the observational
predictions for the primordial perturbations generated by inflation in
these models; in Subsection \ref{inflaton_decay} we present the main
decay channels at the end of inflation; and in Subsection
\ref{reheating} we obtain the reheating temperature after inflation.
In Subsection \ref{compatibility_with_inflation} we consider whether
these inflationary models are compatible with the mass bound presented
in Subsection \ref{BosonMassBound}: in the case of MDSM$_{3}$ we find
compatibility; in the cases of MDSM$_{1}$ and MDSM$_{2}$ we find
incompatibility.  In Subsection \ref{previous_model_building}, we
discuss the relation between this inflationary model, and some
previous models that have been discussed in the literature.

In Section \ref{DarkLeptoViaDecay} we consider the possibility that
the observed dark matter density and matter/anti-matter asymmetry were
both produced directly by the decay of the inflaton into heavy
neutrinos at the end of inflation.  In this scenario, one of the heavy
neutrinos is the dark matter: we compute what its mass should be, find
that it is automatically a non-thermal {\it cold} dark matter
candidate, and show that if this scenario is right, one of the three
light neutrinos must be essentially massless.  Then we discuss the
possibility that one of the other heavy neutrino species, directly
produced during inflaton decay, can generate the cosmological
matter/anti-matter asymmetry through its CP violating decay; this is
(non-thermal) leptogenesis \cite{Fukugita:1986hr}.  Again, we find
this picture seems to be compatible with MDSM$_{3}$, but in tension or
outright conflict with MDSM$_{1}$ and MDSM$_{2}$.

In Section \ref{DarkLeptoViaOther} we draw the reader's attention to
other routes by which dark matter and the cosmological
matter/anti-matter asymmetry may be produced in this model.  Since
these alternative scenarios do not involve an early epoch of inflation
driven by $\psi_{\parallel}$, the tension with MDSM$_{1}$ and
MDSM$_{2}$ is relieved.  In Section \ref{MDSM3_is_nice} we draw
together and emphasize the particularly compelling features of
MDSM$_{3}$, including several features that were not mentioned earlier
in the paper.  Finally, in Section \ref{discuss}, we discuss some
directions for future work.

%%%%%%%%%%%%%%%%%%%%%%%%%%%%%%%%%%%%%%%%%%%%%%%%%%%
\section{The base model and 3 minimal extensions}
\label{MDSM}

The three models we will be interested in (MDSM$_{1}$, MDSM$_{2}$ and
MDSM$_{3}$) are all slight variants of a common (but non-viable) base
model MDSM$_{0}$.  It is convenient to first specify MDSM$_{0}$, and
then describe the 3 variants in turn.

%%%%%%%%%%%%%%%%%%%%%%%%
\subsection{MDSM$_{0}$ (the base model)}
\label{MDSM0}

In brief, the base model MDSM$_{0}$ is the standard model plus 3
right-handed neutrinos ({\it i.e.}\ the ``$\nu$MSM''
\cite{Asaka:2005an, Asaka:2005pn}), augmented by the additional
constraint of classical conformal invariance to eliminate all
dimensionful couplings.

The base model has the same gauge group and field content as the
minimal $SU(3)_{C}\times SU(2)_{L}\times U(1)_{Y}$ standard model,
including 3 gauge-singlet right-handed neutrinos (one per generation).
In other words, the fields and representations in MDSM$_{0}$ are
summarized by the following table:
\begin{equation}
  \label{SM_table}
  \begin{array}{c|c|c|c} 
    & SU(3)_{C} & SU(2)_{L} & U(1)_{Y} \\
    \hline
    q_{L} & 3 & 2 & +1/6 \\
    \hline
    u_{R} & 3 & 1 & +2/3 \\
    \hline
    d_{R} & 3 & 1 & -1/3 \\
    \hline
    l_{L} & 1 & 2 & -1/2 \\
    \hline
    \nu_{R} & 1 & 1 & 0 \\
    \hline
    e_{R} & 1 & 1 & -1 \\
    \hline
    h & 1 & 2 & +1/2
  \end{array}
\end{equation}
Here $q_{L}$ is the left-handed quark doublet, $u_{R}$ and
$d_{R}$ are the right-handed quark singlets, $l_{L}$ is the
left-handed lepton doublet, $\nu_{R}$ and $e_{R}$ are the right-handed
lepton singlets, and $h$ is the scalar Higgs doublet.  As usual, the
fermion (quark and lepton) fields all come in 3 generations that are
identical, apart from the values of their Yukawa couplings.  Our base
model is now obtained by writing down the most general renormalizable
Lagrangian built from these ingredients, with classical conformal
invariance ({\it i.e.}\ dimensionless couplings)
\begin{eqnarray}
  {\cal L}_{0}&=&
  -(1/4)B_{\mu\nu}^{2}\!-\!(1/4)W_{\mu\nu}^{(a)2}
  \!-\!(1/4)G_{\mu\nu}^{(a)2}\!+\!(D_{\mu}h)^{\dagger}(D^{\mu}h)
  \!-\!\lambda_{h}(h^{\dagger}h)^{2} \nonumber\\
  &&+i\bar{q}_{L}D\!\!\!\!/\,q_{L}
  \!+\!i\bar{u}_{R}D\!\!\!\!/\,u_{R}
  \!+\!i\bar{d}_{R}D\!\!\!\!/\,d_{R}
  \!+\!i\bar{l}_{L}D\!\!\!\!/\,l_{L}
  \!+\!i\bar{\nu}_{R}D\!\!\!\!/\,\nu_{R}
  \!+\!i\bar{e}_{R}D\!\!\!\!/\,e_{R} \nonumber\\
  &&
  -\bar{q}_{L}Y_{u}^{\dagger}u_{R}\tilde{h}
  \!-\!\bar{q}_{L}Y_{d}^{\dagger}d_{R}h
  \!-\!\bar{l}_{L}Y_{\nu}^{\dagger}\nu_{R}\tilde{h}
  \!-\!\bar{l}_{L}Y_{e}^{\dagger}e_{R}h+h.c. 
\end{eqnarray}
where $Y_{u}$, $Y_{d}$, $Y_{\nu}$ and $Y_{e}$ are the four $3\times3$
Yukawa matrices, whose indices run over the 3 fermion generations; and
here and throughout the rest of the paper, we will use the following
notation for charge conjugate fields:
\begin{equation}
  \tilde{h}\equiv i\sigma^{2}h^{\ast},\qquad\qquad
  \nu_{c}\equiv i\gamma^{2}\nu_{R}^{\ast}. 
\end{equation}
We will often work in ``unitary gauge,'' in which $h$ is written in
the form
\begin{equation}
  h=\frac{1}{\sqrt{2}}\left(\begin{array}{c}
      0 \\ v_{h}+\rho_{h} \end{array}\right)
\end{equation}
where $v_{h}$ is a constant (the VEV) and $\rho_{h}$ is a field (the
displacement from the VEV).

The classical conformal invariance has the effect of removing two key
terms that are otherwise present and play a phenomenologically
important role in the $\nu$MSM Lagrangian: (i) first, the terms
$\frac{1}{2}\bar{\nu}_{c} M_{m}^{\dagger}\nu_{R}+h.c.$ that give rise
to see-saw masses for the neutrinos; and (ii) second, the term
$m_{h}^{2}h^{\dagger}h$ that generates spontaneous symmetry breaking
at tree level.

Note that MDSM$_{0}$ is not viable by itself. As we shall review in
Subsection \ref{BosonMassBound}, because the top quark is so heavy
(relative to the $W$ and $Z$ bosons), radiative corrections
destabilize the theory: the one-loop effective Higgs potential is
unbounded below \cite{Sher:1988mj}.

Let us now present three minimal extensions of this base model that
retain its renormalizability and classical conformal invariance, but
re-introduce see-saw masses for the neutrinos, and also achieve
successful spontaneous electroweak symmetry breaking via the
Coleman-Weinberg mechanism.

%%%%%%%%%%%%%%%%%%%%%%%%%%%
\subsection{MDSM$_{1}$}
\label{MDSM1}

MDSM$_{1}$ is obtained by adding a single real scalar field $\varphi$,
which is a gauge singlet under $SU(3)_{C}\times SU(2)_{L}\times
U(1)_{Y}$, and again writing down the most general renormalizable
Lagrangian, with classical conformal invariance ({\it i.e.}\
dimensionless couplings):
\begin{equation}
  {\cal L}_{1}={\cal L}_{0}
  \!+\!\frac{1}{2}(\partial\varphi)^{2}
  \!-\!\frac{\lambda_{\varphi}}{4}\varphi^{4}
  \!-\!\lambda_{m}(h^{\dagger}h)\varphi^{2}
  \!-\!\frac{1}{2}\frac{\varphi}{\sqrt{2}}(
  \bar{\nu}_{c}Y_{m}^{\dagger}\nu_{R}+h.c.)
\end{equation}
where $Y_{m}=Y_{m}^{T}$.  As we did with $h$, we split $\varphi$ into
a constant $v_{\varphi}$ (the VEV) and a field $\rho_{\varphi}$ (the
displacement from the VEV):
\begin{equation}
  \varphi=v_{\varphi}+\rho_{\varphi}.
\end{equation}

%%%%%%%%%%%%%%%%%%%%%%%%%%%%%%
\subsection{MDSM$_{2}$}
\label{MDSM2}

MDSM$_{2}$ is the same as MDSM$_{1}$, except now the gauge singlet
scalar field $\varphi$ is complex rather than real, so the most
general renormalizable Lagrangian with classical conformal invariance
is
\begin{equation}
  {\cal L}_{2}={\cal L}_{0}
  \!+\!|\partial\varphi|^{2}\!-\!\lambda_{\varphi}|\varphi|^{4}
  \!-\!2\lambda_{m}(h^{\dagger}h)|\varphi|^{2}
  \!-\!\frac{1}{2}(\varphi\bar{\nu}_{c}Y_{m}^{\dagger}\nu_{R}\!+\!h.c.)
\end{equation}
where again $Y_{m}=Y_{m}^{T}$.  We now write $\varphi$ as
\begin{equation}
  \varphi=\frac{1}{\sqrt{2}}(v_{\varphi}+\rho_{\varphi}){\rm
    e}^{ia/v_{\varphi}}.
\end{equation}
Note the presence in this model of the additional real field $a$,
which measures the complex phase of $\varphi$.  This field, called the
Majoron \cite{Chikashige:1980ui}, is a Goldstone boson at tree level.
Refs.~\cite{Meissner:2008gj, Latosinski:2010qm} argue that $a$ obtains
a mass via quantum effects, and from a phenomenological standpoint is
very much like an axion\footnote{This offers the possibility that 
the strong CP problem may be solved in MDSM$_{2}$.  As far as we are
aware, it is not solved in MDSM$_{1}$ or MDSM$_{3}$.}.

%%%%%%%%%%%%%%%%%%%%%%%%%%%%%%
\subsection{MDSM$_{3}$}
\label{MDSM3}

MDSM$_{3}$ is similar to MDSM$_{2}$, except now we also add a new
$U(1)_{X}$ gauge symmetry, carried by a new gauge boson $C_{\mu}$.
Without loss of generality (see Subsection \ref{anomalies}) this new
charge may be taken to be nothing but baryon number minus lepton
number ($X=B-L$); and in order to preserve see-saw neutrino masses,
the scalar field $\varphi$ must couple to this new symmetry with
charge $+2$.  The fields and their representations are now summarized
by the following table
\begin{equation}
  \label{MDSM3_table}
  \begin{array}{c|c|c|c|c} 
    & SU(3)_{C} & SU(2)_{L} & U(1)_{Y} & U(1)_{X} \\
    \hline
    q_{L} & 3 & 2 & +1/6 & +1/3 \\
    \hline
    u_{R} & 3 & 1 & +2/3 & +1/3 \\
    \hline
    d_{R} & 3 & 1 & -1/3 & +1/3 \\
    \hline
    l_{L} & 1 & 2 & -1/2 & -1 \\
    \hline
    \nu_{R} & 1 & 1 & 0 & -1 \\
    \hline
    e_{R} & 1 & 1 & -1 & -1 \\
    \hline
    h & 1 & 2 & +1/2 & 0 \\
    \hline
    \varphi & 1 & 1 & 0 & +2
  \end{array}
\end{equation}
and the most general renormalizable Lagrangian with classical
conformal symmetry built from these ingredients is
\begin{equation}
  {\cal L}_{3}\!=\!{\cal L}_{0}\!+\!|D_{\mu}\varphi|^{2}
  \!-\!\lambda_{\varphi}|\varphi|^{4}\!-\!2\lambda_{m}
  (h^{\dagger}h)|\varphi|^{2}\!-\!\frac{1}{2}
  (\varphi\bar{\nu}_{c}Y_{m}^{\dagger}\nu_{R}\!+\!h.c.)
  \!-\!\frac{1}{4}C_{\mu\nu}^{2}-\frac{\kappa}{2}B_{\mu\nu}C^{\mu\nu}
\end{equation}
where $Y_{m}=Y_{m}^{T}$.  We fix the final $U(1)_{X}$ gauge freedom by 
setting the imaginary part of $\varphi$ to zero, and then write it as
\begin{equation}
  \varphi=\frac{1}{\sqrt{2}}(v_{\varphi}+\rho_{\varphi}).
\end{equation}

%%%%%%%%%%%%%%%%%%%%%%%%%%%%%%%%%%%%%%%%%%%%%%%%%
\section{Tree-level masses}
\label{tree_level_masses}

In this section, we present the tree-level mass spectra in MDSM$_{1}$,
MDSM$_{2}$ and MDSM$_{3}$.  These will be needed in our subsequent
analysis.

%%%%%%%%%%%%%%%%%%%%%%%%%%%%%%
\subsection{Scalar masses}
\label{tree_level_scalar_masses}

\begin{figure}
  \begin{center}
    \includegraphics[width=4.0in]{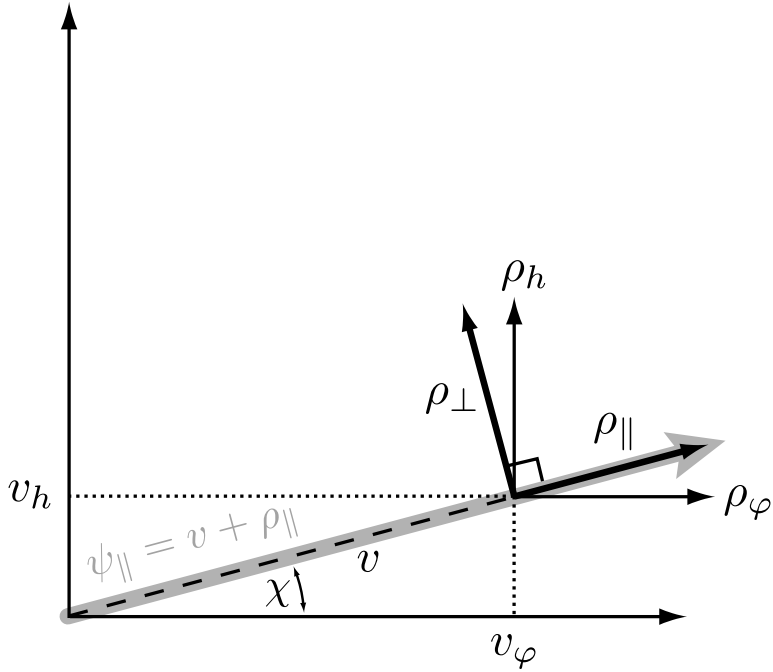}
  \end{center}
  \caption{This figure depicts the relationship (in the space of
    scalar fields) between the scalar VEV (with coordinate 
    $\{v_{\varphi}, v_{h}\}$), the inflaton direction
    $\psi_{\parallel}$, and the scalar mass eigenstates
    ($\rho_{\parallel}$ and $\rho_{\perp}$).}
 \label{vevFig}
\end{figure}

MDSM$_{1}$ and MDSM$_{3}$ contain two physical scalar fields,
$\rho_{h}$ and $\rho_{\varphi}$, that give rise to two physical mass
eigenstates $\rho_{\parallel}$ and $\rho_{\perp}$; MDSM$_{2}$ has, in
addition, a third scalar field $a$ that is a massless Goldstone boson
at the classical level, but obtains a mass via quantum effects and is
similar the usual QCD axion.  Let us start by studying the
$\rho_{\parallel}$ and $\rho_{\perp}$ bosons, since they are present
in all 3 models.  In all 3 models, the tree-level scalar potential
becomes
\begin{equation}
  \label{scalar_potential_tree_level}
  \frac{\lambda_{\varphi}}{4}(v_{\varphi}+\rho_{\varphi})^{4}
  +\frac{\lambda_{m}}{2}(v_{\varphi}+\rho_{\varphi})^{2}
  (v_{h}+\rho_{h})^{2}+\frac{\lambda_{h}}{4}(v_{h}+\rho_{h})^{4}.
\end{equation}
In describing the VEVs, it will be convenient to switch from 
cartesian to polar coordinates by introducing the notation
(see Fig.~\ref{vevFig})
\begin{equation}
  \left(\begin{array}{c} v_{\varphi} \\ v_{h} \end{array}\right)
  =\left(\begin{array}{c} v\,{\rm cos}\,\chi \\
      v\,{\rm sin}\,\chi \end{array}\right).
\end{equation}
Since $\rho_{\varphi}$ and $\rho_{h}$ represent displacements from the
VEV, the terms in (\ref{scalar_potential_tree_level}) linear in
$\rho_{\varphi}$ and $\rho_{h}$ must vanish, leading to the conditions
\begin{subequations}
  \label{degenerate_ray}
  \begin{eqnarray}
    \lambda_{\varphi}{\rm cos}^{2}\chi+\lambda_{m}
    {\rm sin}^{2}\chi&=&0 \\
    \lambda_{h}{\rm sin}^{2}\chi+\lambda_{m}{\rm cos}^{2}\chi &=&0.
  \end{eqnarray}
\end{subequations}
When these conditions are satisfied, the quadratic terms in 
(\ref{scalar_potential_tree_level}) reduce to
\begin{equation}
  \frac{1}{2}(m_{\parallel}^{2}\rho_{\parallel}^{2}
  +m_{\perp}^{2}\rho_{\perp}^{2})
\end{equation}
where the mass eigenstates and eigenvalues are (see Fig.~\ref{vevFig})
\begin{equation}
  \left(\begin{array}{c} \rho_{\parallel} \\ 
      \rho_{\perp} \end{array}\right)
  =\left(\begin{array}{cc} {\rm cos}\chi & {\rm sin}\chi \\
      -{\rm sin}\chi & {\rm cos}\chi \end{array}\right)
  \left(\begin{array}{c} \rho_{\varphi} \\
      \rho_{h} \end{array}\right)\qquad\quad
  \begin{array}{lll}
    m_{\parallel} & = & 0 \\
    m_{\perp} & = & (-2\lambda_{m})^{1/2}v
  \end{array}
\end{equation}
Finally, the axion-like field $a$ in MDSM$_{2}$ is a Goldstone boson
at the classical level, with tree-level mass $m_{a}=0$.

%%%%%%%%%%%%%%%%%%%%%%%%%%%%%%
\subsection{Fermion masses}
\label{tree_level_fermion_masses}

All 3 models have the same fermion content, and the same 
expressions for the fermion masses at tree level.  To see this,
note that in each model, when we expand around the scalar
VEVs, the fermion mass terms are
\begin{eqnarray}
  \label{fermion_mass_terms}
  \!\!&\!\!\!\!&\!\!-\frac{v_{h}}{\sqrt{2}}[
  (\bar{u}_{L}Y_{u}^{\dagger}u_{R}\!+\!\bar{u}_{R}Y_{u}^{}u_{L})\!+\!
  (\bar{d}_{L}Y_{d}^{\dagger}d_{R}\!+\!\bar{d}_{R}Y_{d}^{}d_{L})\!+\!
  (\bar{e}_{L}Y_{e}^{\dagger}e_{R}\!+\!\bar{e}_{R}Y_{e}^{}e_{L})]
  \nonumber\\
  \!\!&\!\!\!\!&\!\!-\frac{v_{h}}{\sqrt{2}}
  (\bar{\nu}_{L}Y_{\nu}^{\dagger}\nu_{R}+\bar{\nu}_{R}Y_{\nu}^{}\nu_{L})-
  \frac{1}{2}\frac{v_{\varphi}}{\sqrt{2}}(\bar{\nu}_{c}Y_{m}^{\dagger}\nu_{R}
  +\bar{\nu}_{R}Y_{m}^{}\nu_{c}).
\end{eqnarray}
To obtain the masses corresponding to the terms on the first line of
(\ref{fermion_mass_terms}), we perform a singular value decomposition,
$Y_{\alpha}=U_{\alpha}y_{\alpha}V_{\alpha}^{\ast}$, where $U_{\alpha}$
and $V_{\alpha}$ are unitary matrices, and $y_{\alpha}$ is diagonal
with real, non-negative eigenvalues $y_{\alpha}^{(1)}\leq
y_{\alpha}^{(2)} \leq y_{\alpha}^{(3)}$.  Then the masses of the
up-type, down-type, and electron-type fermions are
\begin{equation}
  m_{u}^{(i)}=\frac{v_{h}}{\sqrt{2}}y_{u}^{(i)},\quad
  m_{d}^{(i)}=\frac{v_{h}}{\sqrt{2}}y_{d}^{(i)},\quad
  m_{e}^{(i)}=\frac{v_{h}}{\sqrt{2}}y_{e}^{(i)}.
\end{equation}
To obtain the masses corresponding to the terms on the second line
of (\ref{fermion_mass_terms}) -- {\it i.e.}\ the neutrino masses -- let
us start by collecting all of these terms together into a single term
of the form
\begin{equation}
  -\frac{1}{2}\nu^{T}\gamma^{0}\gamma^{2}M\nu+h.c.
\end{equation}
Here $\nu$ is a 6 component vector consisting of all 6 left-handed
neutrinos (3 $\nu_{L}$'s and 3 $\nu_{c}$'s), and $M$ is the
corresponding $6\times 6$ neutrino mass matrix:
\begin{equation}
  \nu\equiv\left(\begin{array}{c} \nu_{L} \\ \nu_{c} \end{array}\right)
  \qquad
  M\equiv\left(\begin{array}{cc} 0 & M_{\nu}^{T} \\
      M_{\nu} & M_{m} \end{array}\right)
\end{equation}
where the $3\times3$ matrices $M_{\nu}$ and $M_{m}$ are given by
\begin{equation}
  M_{\nu}\equiv\frac{v_{h}}{\sqrt{2}}Y_{\nu}\qquad
  M_{m}\equiv\frac{v_{\varphi}}{\sqrt{2}}Y_{m}.
\end{equation}
Now if we regard the $3\times3$ matrix
\begin{equation}
  r\equiv M_{m}^{-1}M_{\nu}^{}
\end{equation}
as ``small'' (as it is in the see-saw mechanism), and block
diagonalize the matrix $M$ through 2nd order in $r$, the result
is
\begin{equation}
  M=S^{T}\left(\begin{array}{cc} M_{n} & 0 \\
      0 & M_{N} \end{array}\right)S
\end{equation}
where $M_{n}$ and $M_{N}$ are the $3\times3$ mass matrices
for the light neutrinos ($n$) and the heavy neutrinos
($N$) respectively
\begin{equation}
  M_{n}^{}=-r^{T}M_{m}^{}r,\qquad
  M_{N}^{}=M_{m}^{}+\frac{1}{2}rr^{T}M_{m}^{}+\frac{1}{2}M_{m}^{}rr^{T}
\end{equation}
and the orthogonal matrix that achieves the block diagonalization is
\begin{equation}
  S=\left(\begin{array}{cc}
      1-(1/2)r^{T}r & -r^{T} \\
      r & 1-(1/2)rr^{T} \end{array}\right).
\end{equation}
Thus, the relation between the original basis ($\nu_{L}$, $\nu_{c}$)
and the block diagonal basis ($n$, $N$) for the neutrino states is
\begin{equation}
  \left(\begin{array}{c} n \\ N \end{array}\right)=
  S\left(\begin{array}{c} \nu_{L} \\ \nu_{c} \end{array}\right)
  \qquad
  \left(\begin{array}{c} \nu_{L} \\ \nu_{c} \end{array}\right)=
  S^{T}\left(\begin{array}{c} n \\ N \end{array}\right).
\end{equation}

%%%%%%%%%%%%%%%%%%%%%%%%%%%%%%
\subsection{Vector masses}
\label{tree_level_vector_masses}

In MDSM$_{1}$ and MDSM$_{2}$, the gauge bosons are exactly the same as
in the ordinary standard model: associated with the unbroken factor
$SU(3)_{C}$ there are 8 massless gluons; and associated with the
spontaneously broken factor $SU(2)_{L}\times U(1)_{Y}$ there are 4
gauge bosons ($W_{\mu}^{(1)}$, $W_{\mu}^{(2)}$, $W_{\mu}^{(3)}$,
$B_{\mu}$) that, after spontaneous symmetry breaking, become 3 massive
vector bosons ($W_{\mu}^{+}$, $W_{\mu}^{-}$, $Z_{\mu}$) and one
massless gauge boson (the photon, $\gamma_{\mu}$, which couples to
electric charge: $Q=Y+T_{3}$):
\begin{equation}
  \begin{array}{llllll}
    W_{\mu}^{\pm} & = & \;(W_{\mu}^{(1)}
    \pm i W_{\mu}^{(2)})/\sqrt{2} & \qquad m_{W}^{} & = & \;g_{w}v_{h}/2 \\
    Z_{\mu} & = & \;{\rm cos}\,\theta_{W} W_{\mu}^{(3)}
    -{\rm sin}\,\theta_{W} B_{\mu} & \qquad m_{Z}^{} & = & 
    \;g_{w}v_{h}/2{\rm cos}\,\theta_{W} \\
    \gamma_{\mu} & = & \;{\rm sin}\,\theta_{W} W_{\mu}^{(3)}
    +{\rm cos}\,\theta_{W} B_{\mu} & \qquad m_{\gamma} & = & \;0
  \end{array}
\end{equation}
where $\theta_{W}$ is the Weinberg angle:
\begin{equation}
  {\rm cos}\,\theta_{W}=\frac{g_{w}^{}}{\sqrt{g_{w}^{2}+g_{y}^{2}}}
  \qquad
  {\rm sin}\,\theta_{W}=\frac{g_{y}^{}}{\sqrt{g_{w}^{2}+g_{y}^{2}}}.
\end{equation}

In MDSM$_{3}$, the situation is slightly more complicated.  Associated
with the unbroken factor $SU(3)_{C}$ there are again 8 massless
gluons.  Associated with the spontaneously broken factor $SU(2)_{L}
\times U(1)_{Y}\times U(1)_{X}$ there are 5 gauge bosons
($W_{\mu}^{(1)}$, $W_{\mu}^{(2)}$, $W_{\mu}^{(3)}$, $B_{\mu}$,
$C_{\mu}$) that, after spontaneous symmetry breaking, become 4 massive
vector bosons ($W_{\mu}^{(+)}$, $W_{\mu}^{(-)}$, $Z_{\mu}$,
$Z_{\mu}'$) and one massless gauge boson (the photon, $\gamma_{\mu}$,
which again couples to electric charge $Q=Y+T_{3}$).  Let us write
these 5 mass eigenstates and eigenvalues more explicitly.  The two
electrically charged states ($W_{\mu}^{+}$ and $W_{\mu}^{-}$) are
exactly the same as before:
\begin{equation}
  W_{\mu}^{\pm}=\frac{1}{\sqrt{2}}(W_{\mu}^{(1)}\pm i W_{\mu}^{(2)}),
  \qquad m_{W}^{}=g_{w}v_{h}/2
\end{equation}
while the photon is given by 
\begin{equation}
  \gamma_{\mu}={\rm sin}\,\theta_{W}^{}\,W_{\mu}^{(3)}
  +{\rm cos}\,\theta_{W}^{}[B_{\mu}+\kappa C_{\mu}],
  \qquad m_{\gamma}=0.
\end{equation}
The remaining two electrically neutral bosons ($Z_{\mu}=Z_{\mu}^{-}$
and $Z_{\mu}'=Z_{\mu}^{+}$) are given by the relatively complicated
formulae:
\begin{equation}
  Z_{\mu}^{\pm}=\frac{1}{d_{\pm}}\left[a_{\pm}W_{\mu}^{(3)}
    +b_{\pm}B_{\mu}+c_{\pm}C_{\mu}\right],\qquad m_{\pm}^{2}=(1/4)v_{h}^{2}\lambda_{\pm}
\end{equation}
where we have defined the constants $p=4 g_{x}(v_{\varphi}/v_{h})$ and:
\begin{subequations}
  \begin{eqnarray}
   &\lambda_{\pm}=\frac{1}{2}\left\{\frac{p^{2}+g_{y}^{2}}{1-\kappa^{2}}
      +g_{w}^{2}\pm\sqrt{\left[\frac{p^{2}-g_{y}^{2}}
          {1-\kappa^{2}}-g_{w}^{2}\right]^{2}+\frac{4\kappa^{2}p^{2}g_{y}^{2}}
        {(1-\kappa^{2})^{2}}}\right\}& \\
    &a_{\pm}=\frac{g_{y}g_{w}}{\lambda_{\pm}^{}-g_{w}^{2}},\quad 
    b_{\pm}=\frac{\kappa^{2}\lambda_{\pm}^{}}{\lambda_{\pm}^{}-p^{2}}-1,\quad
    c_{\pm}=\frac{\kappa p^{2}}{\lambda_{\pm}-p^{2}}& \\
    &d_{\pm}=\sqrt{a_{\pm}^{2}+c_{\pm}^{2}+(1-\kappa^{2})}&
  \end{eqnarray}
\end{subequations}
The linear transformation that relates ($W_{\mu}^{(3)}$, $B_{\mu}$,
$C_{\mu}$) to ($\gamma_{\mu}$, $Z_{\mu}$, $Z_{\mu}'$) is defined by
the requirement that, not only do the mass terms become diagonal, but
kinetic terms also become diagonal and canonically normalized; it is
only an orthogonal transformation when $\kappa=0$.  When the ratio
$v_{\varphi}/v_{h}$ is large (as it will be for us) the expressions
simplify, at leading order, to:
\begin{equation}
  \begin{array}{llllll}
    Z_{\mu}&\approx&{\rm cos}\,\theta_{W} W_{\mu}^{(3)}
    \!-\!{\rm sin}\,\theta_{W}[B_{\mu}\!+\!\kappa C_{\mu}],&\quad
    m_{Z}^{}&\approx&g_{w} v_{h}/2{\rm cos}\,\theta_{W} \\
    Z_{\mu}'&\approx&\sqrt{1\!-\!\kappa^{2}}C_{\mu},&\quad
    m_{Z'}^{}&\approx&2(1\!-\!\kappa^{2})^{-1/2}g_{x}v_{\varphi}
  \end{array}
\end{equation}

%%%%%%%%%%%%%%%%%%%%%%%%%%%%%%%%%%%%%%%%%%%%%%%%%%%%%%%
\section{Symmetry breaking via the Coleman-Weinberg Mechanism}
\label{ColemanWeinberg}

%%%%%%%%%%%%%%%%%%%%%%%%%%%%%%
\subsection{The Gildener-Weinberg formalism}
\label{GildenerWeinberg}

An elegant and useful formalism for analyzing Coleman-Weinberg
symmetry breaking \cite{Coleman:1973jx} in theories with arbitrary
scalar field content is developed by E.~Gildener and S.~Weinberg in
Ref.~\cite{Gildener:1976ih}.  For a detailed justification of the
following calculations, we refer the reader to
Ref.~\cite{Gildener:1976ih}.  In this section, we just summarize the
essential results.  The first step is to run the renormalization group
scale to a special scale $\Lambda$ at which the tree-level scalar
potential has a degenerate valley of minima along a ray extending out
from the origin in field space: in our models, this is equivalent to
working at the renormalization scale $\Lambda$ at which the tree-level
parameters in our effective potential satisfy
Eq.~(\ref{degenerate_ray}), with $\lambda_{m}<0$.  At this scale,
define two constants, $A$ and $B$:
\begin{subequations}
  \label{def_AB}
  \begin{eqnarray}
    \label{def_A}
    A&\equiv&\frac{1}{64\pi^{2}}\left[
      \sum_{s}\frac{m_{s}^{4}}{v^{4}}{\rm ln}\frac{m_{s}^{2}}{v^{2}}
      \!-\!4\sum_{f}\zeta_{f}\frac{m_{f}^{4}}{v^{4}}{\rm ln}\frac{m_{f}^{2}}{v^{2}}
      \!+\!3\sum_{v}\frac{m_{v}^{4}}{v^{4}}{\rm ln}\frac{m_{v}^{2}}{v^{2}}
    \right] \\
    \label{def_B}
    B&\equiv&\frac{1}{64\pi^{2}}\left[
      \sum_{s}\frac{m_{s}^{4}}{v^{4}}\qquad\;
      -\!4\sum_{f}\zeta_{f}\frac{m_{f}^{4}}{v^{4}}\qquad\;
      +\!3\sum_{v}\frac{m_{v}^{4}}{v^{4}}\qquad\;\;\!\right]
  \end{eqnarray}
\end{subequations}
where, in each equation, the 3 sums are over all tree-level scalar,
fermion, and vector masses, respectively; and $\zeta_{f}=1$ for Dirac
fermions and $1/2$ for Majorana or Weyl fermions.  These sums
are dominated by largest tree-level masses in the theory; so, 
in MDSM$_{1}$, MDSM$_{2}$ and MDSM$_{3}$, respectively, we have
\begin{subequations}
  \label{B1_B2_B3}
  \begin{eqnarray}
    B_{0}&\approx&\frac{1}{64\pi^{2}v^{4}}\left[
      -12 m_{t}^{4}\!-\!2\sum_{i=1}^{3}m_{N,i}^{4}
      \!+\!6 m_{W}^{4}\!+\!3 m_{Z}^{4}\right] \\
    B_{1}=B_{2}&\approx&\frac{1}{64\pi^{2}v^{4}}\left[
     -12 m_{t}^{4}\!-\!2\sum_{i=1}^{3}m_{N,i}^{4}
      \!+\!6 m_{W}^{4}\!+\!3 m_{Z}^{4}+m_{\perp}^{4}\right] \\
    B_{3}&\approx&\frac{1}{64\pi^{2}v^{4}}\left[
      -12 m_{t}^{4}\!-\!2\sum_{i=1}^{3}m_{N,i}^{4}
      \!+\!6 m_{W}^{4}\!+\!3 m_{Z}^{4}+m_{\perp}^{4}+3 m_{Z'}^{4}\right]. 
 \end{eqnarray}
\end{subequations}
At tree-level, the effective potential had a degenerate valley of
minima along the direction given by angle $\chi$ in field space; but
the one-loop correction gives a radial shape to the potential along
this direction:
\begin{equation}
  \label{CW_potential}
  V(\psi_{\parallel})=\frac{B}{2}v^{4}\left(1-\frac{\psi_{\parallel}^{4}}{v^{4}}+
    4\frac{\psi_{\parallel}^{4}}{v^{4}}{\rm ln}\frac{\psi_{\parallel}}{v}\right).
\end{equation}
This lifts the degeneracy, and picks out a unique minimum a distance
$v$ from the origin in scalar field space (see Fig.~\ref{vevFig}).
For our later cosmological applications, we have added an overall
constant to the potential so that $V=0$ at the minimum
$\psi_{\parallel}=v$ ($\rho_{\parallel}=0$).  In other words, we do
{\it not} solve the cosmological constant problem: we do not explain
why the cosmological constant is small, but merely choose it to be
tiny, in accordance with cosmological observations.  The
$\rho_{\parallel}$ boson was massless at tree level, but at one-loop
obtains a mass
\begin{equation}
  \label{m_parallel_one_loop}
  m_{\parallel}=(8B)^{1/2}v.
\end{equation}
Finally note that the VEV $v$, the renormalization scale $\Lambda$,
and the constants $A$ and $B$ are related to one another via the
constraint
\begin{equation}
  \label{Lambda_v_relation}
  {\rm ln}\frac{v^{2}}{\Lambda^{2}}=-\frac{1}{2}-\frac{A}{B}.
\end{equation}

%%%%%%%%%%%%%%%%%%%%%%%%%%%%%%
\subsection{Lower bounds on new bosonic masses}
\label{BosonMassBound}

We must require $B>0$ so that the one-loop effective potential
(\ref{CW_potential}) is bounded below.  In MDSM$_{0}$, this condition
is not satisfied ($B_{0}<0$).  Thus, as previously mentioned in
Section \ref{intro} and Subsection \ref{MDSM0}, the base model
MDSM$_{0}$ is not viable.  In MDSM$_{1}$ and MDSM$_{2}$, the
requirement $B>0$ implies a lower bound on the $\rho_{\perp}$ mass
\begin{equation}
  \label{BosonMassBound12}
  m_{\perp}>\left[12 m_{t}^{4}-6 m_{W}^{4}-3
    m_{Z}^{4}\right]^{1/4}=318.3~{\rm GeV}.
\end{equation}
In MDSM$_{3}$ it implies, instead, a lower bound on a combination of
the $\rho_{\perp}$ and $Z'$ masses:
\begin{equation}
  \label{BosonMassBound3}
  m_{\perp}^{4}+3 m_{Z'}^{4}>(318.3~{\rm GeV})^{4}.
\end{equation}

%%%%%%%%%%%%%%%%%%%%%%%%%%%%%%
\subsection{Upper bounds on heavy neutrino masses}
\label{NeutrinoMassBound}

One can also interpret the requirement $B>0$ as an upper bound on the
3 heavy neutrino masses $m_{N,i}$; in MDSM$_{1}$ and MDSM$_{2}$
this bound is
\begin{equation}
  \label{NeutrinoMassBound12}
  2\sum_{i}m_{N,i}^{4}<m_{\perp}^{4}-(318.3~{\rm GeV})^{4}
\end{equation}
and in MDSM$_{3}$ it becomes
\begin{equation}
  \label{NeutrinoMassBound3}
  2\sum_{i}m_{N,i}^{4}<3m_{Z'}^{4}+m_{\perp}^{4}-(318.3~{\rm GeV})^{4}.
\end{equation}
%Again, we will see that if we try to embed inflation within these
%models, then the bound (\ref{NeutrinoMassBound12}) will place a very
%tight upper bound on the heavy neutrino masses in MDSM$_{1}$ and
%MDSM$_{2}$; whereas, by contrast, the upper bound 
%(\ref{NeutrinoMassBound3}) on the heavy neutrino masses in 
%MDSM$_{3}$ will be much looser.

%When we try to embed inflation in these models, we will see that 
%the bounds (\ref{BosonMassBound12}, \ref{NeutrinoMassBound12}) on 
%MDSM$_{1}$ and MDSM$_{2}$ will be much more stringent than 
%the bounds (\ref{BosonMassBound3}, \ref{NeutrinoMassBound3}) on 
%MDSM$_{3}$.

%%%%%%%%%%%%%%%%%%%%%%%%%%%%%%%%%%%%%%%%%%%%%%%%%%%%%%
\section{Inflation}
\label{inflation}

%%%%%%%%%%%%%%%%%%%%%%%%%%%%

\subsection{Predicted mass relation}
\label{inflation_mass_relation}

The inflaton candidate in the MDSM is the scalar field
$\psi_{\parallel}$. If $\psi_{\parallel}$ was initially perched near
the point of unbroken symmetry\footnote{One may ask {\it why} the
  field was initially perched near $\psi_{\parallel}=0$.  One
  possibility comes from the ``no-boundary wave function'' (``NBWF'')
  proposal of Hartle and Hawking \cite{Hartle:1983ai}.  Following the
  reasoning in \cite{Hartle:2010dq}, one can show that the
  ``top-down'' prediction of the NBWF for our Coleman-Weinberg-shaped
  potential is, in fact, precisely that the field should have been
  perched near $\psi_{\parallel}=0$ initially.}
($\psi_{\parallel}=0$), standard single-field slow-roll inflation
occurs as it rolls along the $\psi_{\parallel}$ direction toward the
minimum at $\psi_{\parallel}=v$. The shape of the effective potential
along this direction is given by (\ref{CW_potential}). This potential
depends on two parameters ($B$ and $v$), but in order to match the
observed amplitude \cite{Komatsu} of the primordial scalar
perturbations [$\Delta_{{\cal R}}^{2}(k_{\ast})=
(2.43\pm0.1)\times10^{-9}$ at $k_{\ast}=0.002~{\rm Mpc}^{-1}$] these
parameters must be related as shown in the first panel of
Fig.~\ref{InflationFig}. With this constraint, we can regard the
potential (\ref{CW_potential}) as depending on just a single free
parameter $v$.

Recall that $B$ is related to the tree-level mass spectrum of the
theory: see Eqs.~(\ref{def_B}) and (\ref{B1_B2_B3}). In Subsections
\ref{BosonMassBound} and \ref{NeutrinoMassBound}, we used the
requirement $B>0$ to obtain predicted bounds on some of the heaviest
particles in the MDSM. Now, since inflation predicts a specific
relationship between $B$ and $v$, and in particular always predicts
$B\sim10^{-14}$ (see Panel 1 in Fig.~\ref{InflationFig}), we can go
further: the relationship between $B$ and $v$ shown in Panel 1 of
Fig.~\ref{InflationFig} amounts to a specific prediction from
inflation for a mass relation between the heaviest particles in the
MDSM. Is there any way to confirm this prediction (or its relationship
to the other predictions depicted in the subsequent panels of
Fig.~\ref{InflationFig})?

\subsection{Predictions for primordial perturbations}
\label{perturbations}

\begin{figure}
  \begin{center}
    \includegraphics[width=4.5in]{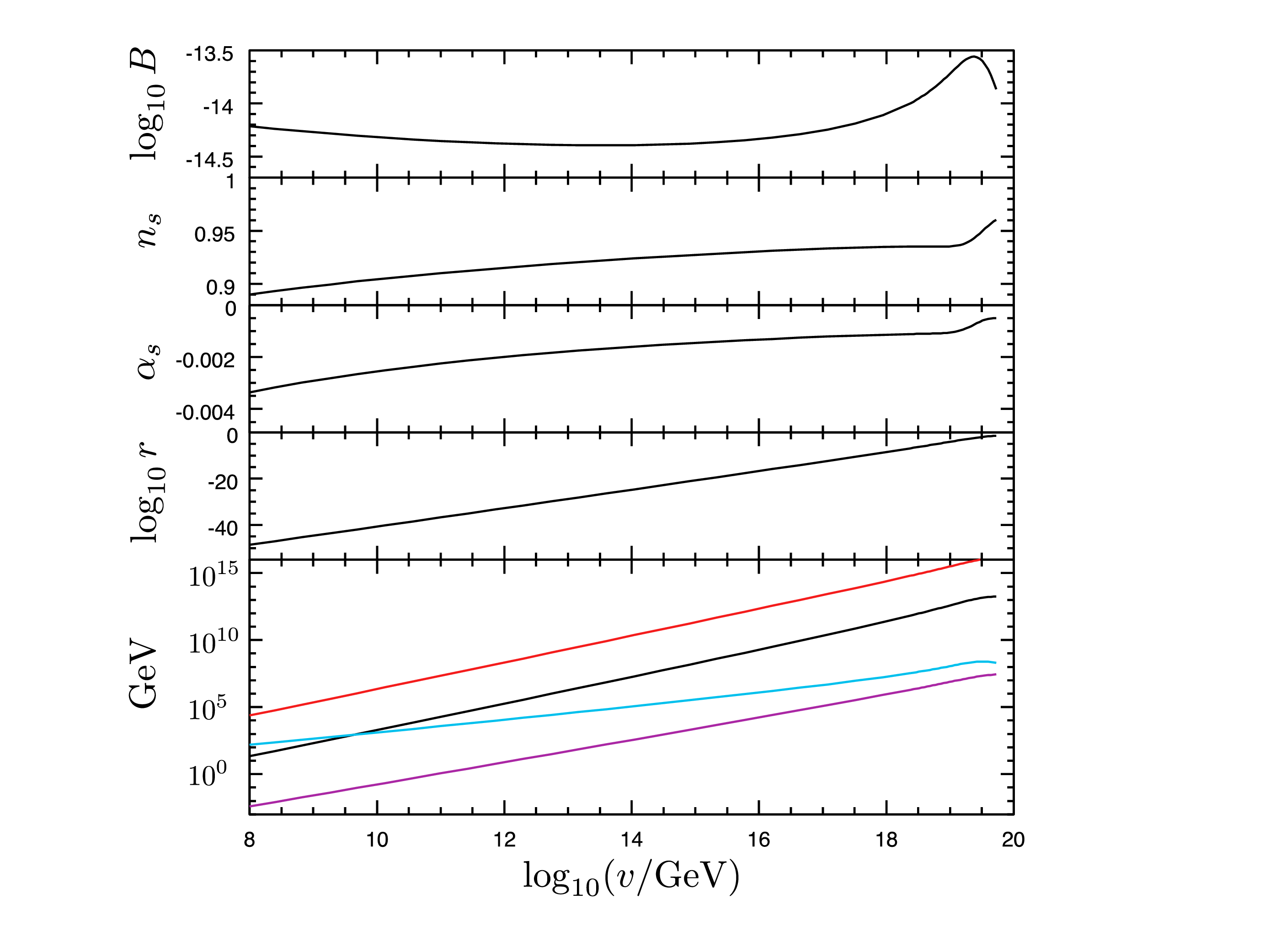}
  \end{center}
  \caption{This figure depicts several predictions of the MDSM
    inflationary scenario, as a function of $v$, the scalar field 
    VEV. The top 4 panels show, respectively, the
    predicted value for $B$, the primordial scalar spectral index
    $n_{s}$, the scalar spectral running $\alpha_{s}
    =dn_{s}/d\,{\rm ln}\,k$, and the primordial tensor-to-scalar
    ratio $r$.  The four curves in the bottom panel show the 
    following: (i) the red curve shows $V_{{\rm max}}^{1/4}$,
    where $V_{{\rm max}}$ is the height of the potential at
    its local maximum at the point of unbroken symmetry;
    (ii) the black curve shows $m_{\parallel}$, the mass of the
    $\rho_{\parallel}$ boson; (iii) the blue curve shows 
    $T_{RH}$, the reheating temperature; and (iv) the purple curve
    shows the dark-matter mass $M_{1}$
    ({\it i.e.}\ the mass of the stable right-handed neutrino which
    gives the correct dark matter abundance).}
 \label{InflationFig}
\end{figure}

Using techniques that are by now standard (see {\it e.g.}\
\cite{Liddle:1994dx}), we compute (numerically) the observable
predictions for the primordial scalar (density) and tensor
(gravitational wave) perturbations predicted by the inflaton potential
(\ref{CW_potential}), as a function of $v$. In particular, in panels 2
through 4 of Fig.~\ref{InflationFig} we plot the predicted scalar
spectral index $n_{s}$, the running of the scalar spectral index
$\alpha_{s}=dn_{s}/d\,{\rm ln}\,k$, and the tensor to scalar ratio $r$
as a function of $v$. Other predictions agree with standard
single-field slow-roll inflation: the scalar perturbations should be
adiabatic (no isocurvature perturbations) and gaussian (negligible
non-gaussianity).

Notice that $n_{s}$ is predicted to be significantly ``red''
($n_{s}<1$), and gets redder as $v$ decreases. Current cosmological
observations place a lower bound on $n_{s}$ which therefore implies an
observational lower bound on $v$ if $\psi_{\parallel}$ is the
inflaton. In particular, for $r\approx0$ (appropriate for the low $v$
regime), observations currently favor $n_{s}\approx0.96$, with a lower
bound $n_{s}\gtrsim0.93$ (at $2\sigma$) or $n_{s}\gtrsim0.92$ (at
$3\sigma$) \cite{Komatsu}. From Panel 2 in Fig.~\ref{InflationFig} we
see that this translates into a lower bound $v\gtrsim 10^{13}~{\rm
  GeV}$ or $v\gtrsim 10^{11}~{\rm GeV}$, respectively. Forthcoming
data from the Planck satellite should significantly tighten the
constraints on $n_{s}$, leading to significantly tighter constraints
on $v$. The observable $\alpha_{s}$ is predicted to be negative, but
with sufficiently small absolute value that its deviation will be
difficult (though not impossible) to detect \cite{Takada}. The
observable $r$ is only large enough to be detected when
$v>10^{19}~{\rm GeV}$.

%%%%%%%%%%%%%%%%%%%%%%%%%%%%%%
\subsection{Inflaton decay} 
\label{inflaton_decay}

At the end of inflation, the energy of the universe is initially
stored in an oscillating condensate of $\rho_{\parallel}$ bosons.  In
this subsection, we will present some of the main decay rates by which
the $\rho_{\parallel}$ decays to other scalar, spinor, and vector
fields.

If the $\rho_{\parallel}$ mass is more than twice the $\rho_{\perp}$
mass, then it can decay to a pair of $\rho_{\perp}$'s via a 3-leg
diagram coming from a term $-(m_{\perp}^{2}/v)\rho_{\parallel}
\rho_{\perp}^{2}$ in the Lagrangian.  The corresponding decay rate is
\begin{equation}
  \Gamma(\rho_{\parallel}\to\rho_{\perp}\rho_{\perp})=
    \frac{m_{\perp}^{4}}{8\pi m_{\parallel}v^{2}}\left(
      1-4\frac{m_{\perp}^{2}}{m_{\parallel}^{2}}\right)^{1/2}. 
\end{equation}
If the $\rho_{\parallel}$ mass is more than twice the mass of fermion
$f$, then it can decay to an $f\bar{f}$ pair via a 3-leg diagram with
vertex factor $-i(m_{f}/v)$.  The decay rate
$\Gamma(\rho_{\parallel}\to f\bar{f})$ depends on whether $f$ is a
quark $q$ (a Dirac fermion with 3 colors), a charged lepton $l$ (a
colorless Dirac fermion), or a neutrino mass eigenstate $\nu$ (a
colorless Majorana fermion); the corresponding rates are
\begin{subequations}
  \begin{eqnarray}
    \Gamma(\rho_{\parallel}\to q\bar{q})&=&\frac{3m_{q}^{2}m_{\parallel}}
    {8\pi v^{2}}\left(1-4\frac{m_{q}^{2}}{m_{\parallel}^{2}}\right)^{3/2} \\
    \Gamma(\rho_{\parallel}\to l\bar{l})&=&\frac{m_{l}^{2}m_{\parallel}}
    {8\pi v^{2}}\left(1-4\frac{m_{l}^{2}}{m_{\parallel}^{2}}\right)^{3/2} \\
    \Gamma(\rho_{\parallel}\to \nu\bar{\nu})&=&\frac{m_{\nu}^{2}m_{\parallel}}
    {16\pi v^{2}}\left(1-4\frac{m_{\nu}^{2}}{m_{\parallel}^{2}}\right)^{3/2}.
  \end{eqnarray}
\end{subequations}
If the $\rho_{\parallel}$ mass is more than twice the mass of vector
$V$, then it can decay into a pair of $V$'s via a 3-leg diagram.  In
particular, it can decay to: (i) a $W^{+}W^{-}$ pair through the
Lagrangian term $-2 (m_{W}^{2}/v) \rho_{\parallel}W^{+}W^{-}$; (ii) a
$ZZ$ pair through the term $-(m_{Z}^{2}/v)\rho_{\parallel}Z^{2}$;
(iii) or (in Model 3 only) a $Z'Z'$ pair through the term
$-(m_{Z'}^{2}/v)\rho_{\parallel}Z^{2}$.  The corresponding rates are
\begin{subequations}
  \begin{eqnarray}
    \Gamma(\rho_{\parallel}\to W^{+}W^{-})&=&
    \frac{m_{\parallel}^{3}}{16\pi v^{2}}\left(1-4
      \frac{m_{W}^{2}}{m_{\parallel}^{2}}\right)^{1/2}
    \left[1-4\frac{m_{W}^{2}}{m_{\parallel}^{2}}
      +12\frac{m_{W}^{4}}{m_{\parallel}^{4}}\right] \\
    \Gamma(\rho_{\parallel}\to ZZ)&=&
    \frac{m_{\parallel}^{3}}{32\pi v^{2}}\left(1-4
      \frac{m_{Z}^{2}}{m_{\parallel}^{2}}\right)^{1/2}
    \left[1-4\frac{m_{Z}^{2}}{m_{\parallel}^{2}}
      +12\frac{m_{Z}^{4}}{m_{\parallel}^{4}}\right] \\
    \Gamma(\rho_{\parallel}\to Z'Z')&=&
    \frac{m_{\parallel}^{3}}{32\pi v^{2}}\left(1-4
      \frac{m_{Z'}^{2}}{m_{\parallel}^{2}}\right)^{1/2}
    \left[1-4\frac{m_{Z'}^{2}}{m_{\parallel}^{2}}
      +12\frac{m_{Z'}^{4}}{m_{\parallel}^{4}}\right]
  \end{eqnarray}
\end{subequations}

%%%%%%%%%%%%%%%%%%%%%%%%%%%%%%
\subsection{Reheating Temperature $T_{RH}$}
\label{reheating}

To compute the reheating temperature $T_{RH}$ (the temperature at the
start of the radiation-dominated epoch), we need to calculate the
total $\rho_{\parallel}$ decay rate, $\Gamma_{\parallel}$.  First
consider the mass of the $\rho_{\parallel}$ boson,
$m_{\parallel}=(8B)^{1/2}v$, the black curve in the top panel of
Fig.~\ref{InflationFig}.  For values of $v$ that are sufficiently
large to be consistent with the observational lower bound on $n_{s}$
(discussed in Subsection \ref{perturbations}), we see that
$m_{\parallel}$ is large relative to the $W$ and $Z$ boson masses; and
then, from inspecting the decay rates presented in the preceding few
subsections, we see that the leading $\rho_{\parallel}$ decay channels
are $\rho_{\parallel}\to W^{+}W^{-}$ and $\rho_{\parallel} \to ZZ$.
Although the contributions of $\Gamma(\rho_{\parallel}\to Z'Z')$ and
$\Gamma(\rho_{\parallel}\to N_{i}N_{i})$ can be also be non-negligible
in some circumstances, they are never large, and neglecting them does
not significantly alter the calculations in this subsection.  Thus, we
can approximate the total $\rho_{\parallel}$ decay rate by the simple
formula
\begin{equation}
  \Gamma_{\parallel}\approx
  \Gamma(\rho_{\parallel}\to W^{+}W^{-})+
  \Gamma(\rho_{\parallel}\to ZZ)\approx 
  \frac{3 m_{\parallel}^{3}}{32\pi v^{2}}.
\end{equation}
From here, we can compute $T_{RH}$ in the standard way, by requiring
that the decay rate $\Gamma_{\parallel}$ should equal the Hubble
expansion rate $H_{RH}$ at the start of the radiation era
\cite{KolbTurner, Mukhanov, Weinberg}:
\begin{equation}
  \Gamma_{\parallel}=\left[\frac{8\pi G}{3}\frac{\pi^{2}}{30}
    g_{\ast}(T_{RH})T_{RH}^{4}\right]^{1/2}.
\end{equation}
We can solve this for $T_{RH}$ to find
\begin{equation}
  T_{RH}=\frac{3}{\pi}\left(\frac{5}{g_{\ast}(T_{RH})}\right)^{1/4}
  B^{3/4}M_{pl}^{1/2}v^{1/2}.
\end{equation}
The bottom panel of Fig.~\ref{InflationFig} shows the reheat
temperature $T_{RH}$ (blue curve), alongside the mass
$m_{\parallel}=(8B)^{1/2}v$ of the $\rho_{\parallel}$ boson (black
curve) and the height of the inflationary hilltop $V_{{\rm
    max}}=(B/2)v^{4}$ [or its 4th root, $V_{{\rm
    max}}^{1/4}=(B/2)^{1/4}v$, which has units of mass] (red curve).

\subsection{Compatibility with MDSM$_{3}$; incompatibility
  with MDSM$_{1}$ and MDSM$_{2}$}
\label{compatibility_with_inflation}

Which of the 3 models (MDSM$_{1}$, MDSM$_{2}$, or MDSM$_{3}$) is
compatible with inflation?  On the one hand, accelerator constraints
require us to have $v_{h}=246~{\rm GeV}$.  On the other hand, we saw
in Subsection \ref{perturbations} that, in order to be compatible with
observational constraints on the primordial scalar spectral index
$n_{s}$, the parameter $v$ must be $\gtrsim10^{11}~{\rm GeV}$.  Thus,
if we want to embed inflation in the MDSM, we are forced into the
regime $\chi\ll1$.  In this regime, the $\rho_{\perp}$ boson is
essentially the ordinary Higgs boson, and its mass is given by
$\rho_{\perp}\approx(2\lambda_{h})^{1/2}v_{h}$.  

In MDSM$_{1}$ and MDSM$_{2}$, the mass bound (\ref{BosonMassBound12})
then says that the mass of the ordinary Higgs boson must be greater
than $318.3~{\rm GeV}$; and when combined with current LHC results,
this lower bound increases to $\sim450~{\rm GeV}$. This lower bound
implies that the dimensionless coupling $\lambda_{h}$ must be
$\gtrsim2$, which is uncomfortably large within perturbation theory.
Furthermore, an ordinary Higgs mass above either $318~{\rm GeV}$ or
$450~{\rm GeV}$ is in tension with indirect limits coming from
precision electroweak data, which prefer a light Higgs, but are rather
broad. Based on these considerations, it seems at this point that
MDSM$_{1}$ and MDSM$_{2}$ are not good candidates to drive inflation;
but it may be too strong to say that this possibility is strictly
ruled out at present\footnote{In fact, in the time since this paper
  first appeared on the arXiv, the ATLAS and CMS collaborations have
  announced the discovery of what is apparently the Higgs boson, with
  a mass near $126~{\rm GeV}$, so that one can now say with much
  greater certainty that MDSM$_{1}$ and MDSM$_{2}$ are ruled out as
  candidates to drive inflation.}.

By contrast, MDSM$_{3}$ is completely compatible with inflation:
the mass bound (\ref{BosonMassBound3}) is readily satisfied, since it
involves {\it both} the Higgs mass $m_{\perp}$ and the $Z'$ mass
$m_{Z'}$.  For example, if forthcoming LHC results determine that the
Higgs mass is $135~{\rm GeV}$, there is no obstruction to setting
$m_{\perp}$ to this value: this corresponds to
$\lambda_{h}\approx0.15$ (comfortably within the range of validity of
perturbation theory) and is compatible with the mass bound
(\ref{BosonMassBound3}) as long as we make the $Z'$ mass sufficiently
large.

%%%%%%%%%%%%%%%%%%%%%%%%
\subsection{Relation to previous inflationary model building}
\label{previous_model_building}

The textbook spontaneous symmetry breaking potential has the form
\begin{equation}
  \label{textbook_V}
  V(\psi)=V_{0}-m^{2}\psi^{2}+\lambda\psi^{4}=
  V_{0}\left[1-\frac{\psi^{2}}{v^{2}}\right]^{2}.
\end{equation}
If one imagines minimally coupling this potential to gravity, and
using it to drive single-field inflation, one finds that it can
generate observationally acceptable primordial perturbations, but only
when the VEV $v$ is larger than the Planck scale
($v\gtrsim10^{19}~{\rm GeV}$); for $v<10^{19}~{\rm GeV}$, the
primordial scalar spectral index becomes unacceptably ``red''
($n_{s}<0.9$). For this reason, the minimal standard model Higgs
doublet $h$, minimally coupled to gravity, is not a viable inflaton
candidate (since its VEV must be $246~{\rm GeV}$).

One way around this problem is to couple the ordinary Higgs boson to
gravity {\it non-minimally} -- {\it i.e.}\ by adding to the Lagrangian
a coupling of the form $\xi (h^{\dagger}h)R$ \cite{Bezrukov:2007ep}.
But one finds that, to match observations, the dimensionless coupling
$\xi$ must be very large ($\xi\sim10^{5}$), which casts doubt on the
desirability and reliability of this solution.

Another solution is to replace the textbook potential
(\ref{textbook_V}) by the Coleman-Weinberg potential
(\ref{CW_potential}), as we have done in this paper. Although the
Coleman-Weinberg potential has played an important role in the history
of inflation since its earliest days \cite{Albrecht:1982wi,
  Linde:1981mu}, as far as we are aware it was Ref.~\cite{Knox:1992iy}
that first emphasized the property that, for us, is essential. Namely,
\cite{Knox:1992iy} emphasized that, because of the particular
shape of the Coleman-Weinberg potential (and, in particular, because
its second derivative $V''(\psi)$ vanishes near the hilltop), it
predicts observationally acceptable primordial perturbations
($0.9<n_{s}<1$) even when the VEV $v$ is orders of magnitude below
the Planck scale. (This is illustrated in the second panel of our
Fig.~\ref{InflationFig}.)

In other words, a nice consequence of starting from a model with no
dimensionful couplings is that the Coleman-Weinberg shape of the
symmetry breaking potential automatically leads to a model that
retains the simplicity, economy and predictivity of single-field
inflation, but nevertheless allows inflation to take place over a
range of field values that is much smaller than the Planck scale. This
is achieved without the need for large dimensionless coupling
constants: {\it e.g.}\ in MDSM$_{3}$, all of the dimensionless
couplings are small.

%%%%%%%%%%%%%%%%%%%%%%%%%%%%%%%%%%%%%%%%%%%%%%%%%%%%%%%
\section{Dark matter and leptogenesis via direct inflaton decay}
\label{DarkLeptoViaDecay}

In this section, we imagine that inflation has just taken place, and
analyze the elegant possibility that the dark matter abundance and
cosmic matter/anti-matter asymmetry can both be accounted for via
direct, non-thermal production of heavy right-handed neutrinos during
inflaton decay.

\subsection{Prediction 1: Mass of the dark matter particle}

Suppose that one of the heavy neutrinos ($N_{1}$) is less than half
the $\rho_{\parallel}$ mass, so that it is directly produced in
$\rho_{\parallel}$ decay after inflation.  In order for the $N_{1}$ to
be produced with the correct abundance to match the presently observed
dark matter density, its mass $M_{1}$ must satisfy the condition [see
{\it e.g.}\ Eq.~(11) in \cite{Gelmini:2006pw}]
\begin{equation}
  1\approx 2\times10^{9}b\,\frac{M_{1}}{m_{\parallel}}
  \frac{T_{RH}}{{\rm GeV}}
\end{equation}
where $b$ is average number of dark matter particles created per
$\rho_{\parallel}$ decay: $b\approx\frac{4}{3}
(M_{1}/m_{\parallel})^{2}$.  Solving for $M_{1}$ we find
\begin{equation}
  \label{dark_matter_mass}
  \frac{M_{1}}{{\rm GeV}}\approx 3\times10^{-6}B^{1/4}
  \left(\frac{v}{{\rm GeV}}\right)^{5/6},
\end{equation}
which is the purple curve in the bottom panel of
Fig.~\ref{InflationFig}.

\subsection{Prediction 2: Dark matter is cold}

When $N_{1}$ is originally produced, it is relativistic, with initial
$\gamma$ factor
\begin{equation}
  \gamma_{i}=\frac{m_{\parallel}}{2M_{1}}
  \approx\frac{2^{1/2}B^{1/4}}{3\times10^{-6}}\left(
    \frac{v}{{\rm GeV}}\right)^{1/6}.
\end{equation}
As the universe cools, it redshifts and eventually becomes
non-relativistic when the temperature of the radiation 
bath (from which it is decoupled) has reached:
\begin{equation}
  T_{NR}\approx\frac{T_{RH}}{\gamma_{i}}\approx
  \frac{9\times10^{-6}}{\pi\sqrt{2}}
  \left(\frac{5}{g_{\ast}(T_{RH})}\right)^{1/4}\left(
    \frac{v}{{\rm GeV}}\right)^{1/3}B^{1/2}M_{pl}^{1/2}{\rm GeV}^{1/2}.
\end{equation}
Thus, as soon as we choose the $N_{1}$ particle to have the
mass $M_{1}$ that is required to obtain the right dark matter
abundance today, we also automatically ensure that this particle 
becomes non-relativistic sufficiently early in cosmic history so that
its free-streaming has a negligible effect on the growth of observed
cosmic structure.  In other words, once we choose the mass $M_{1}$
to get the right dark matter abundance, we automatically obtain 
{\it cold} dark matter, as favored by cosmological observations.

\subsection{Prediction 3: One neutrino is (essentially) massless}

In order to be the dark matter particle, the neutrino $N_{1}$ must
have a lifetime much longer than the current age of the universe (and,
indeed, longer than about $4\times10^{22}$ seconds
\cite{DolgovHansen}).  To achieve this, its effective mixing angle
with the 3 active neutrinos must be extremely small.  From Subsection
\ref{tree_level_scalar_masses}, we see that this implies that one of
the 3 light neutrinos must either be massless, or essentially massless
({\it i.e.}\ with a mass orders of magnitude less than that of the
other two light neutrinos, and much too small to be detected in
practice).  To see that this state of affairs is technically natural,
note that if the Lagrangian were symmetric under $N_{1}\to-N_{1}$, the
$N_{1}$ particle would have zero mixing with light neutrinos, and one
of the light neutrinos would be strictly massless.

\subsection{Leptogenesis via inflaton decay}
\label{LeptogenesisViaDecay}

Now suppose that at least one of the {\it other} two heavy neutrinos
($N_{2}$) is also lighter than half the $\rho_{\parallel}$ mass, so
that it is also directly produced in $\rho_{\parallel}$ decay after
inflation.  The decay of this neutrino produces a primordial lepton
asymmetry; and if this primordial lepton asymmetry is produced early
enough, it can be converted into the observed baryon asymmetry via
sphaleron transitions.  This mechanism for producing the observed
matter/anti-matter asymmetry is known as (non-thermal) leptogensis
\cite{Fukugita:1986hr}.

In order to have successful leptogenesis, we also must first choose
the $N_{2}$ decay rate $\Gamma_{2}$ to satisfy the following
conditions: when the $N_{2}$ decays, its mass must be higher than the
temperature $T_{2}$ of the radiation bath (so that inverse decays are
Boltzmann suppressed), and also $T_{2}$ must be $\gtrsim100~{\rm GeV}$
(so that the sphalerons -- whose energy scale is set by the $W$ and
$Z$ masses -- are still active).  This may be achieved by choosing its
effective mixing angle $\theta_{2}$ with active neutrinos to lie in
the right range.  Once these conditions are satisfied,the condition
for obtaining the correct matter/anti-matter asymmetry becomes
a condition on the mass $M_{2}$ of the $N_{2}$ particle:
\begin{subequations}
  \begin{eqnarray}
    \label{M2_formula}
    \frac{M_{2}}{{\rm GeV}}&\approx&
    2.5\times10^{-4}\left[\frac{(n_{B}/s)}{C(L_{1}-L_{2})(r-\bar{r})}
    \right]^{1/2}B^{3/8}\Big(\frac{v}{{\rm GeV}}\Big)^{5/4} \\
    &\gtrsim&2.5\times10^{-9} B^{3/8}\Big(\frac{v}{{\rm GeV}}\Big)^{5/4}
  \end{eqnarray}
\end{subequations}
Here we are following the notation in \cite{Weinberg}: the $N_{2}$ has
two decay channels: $N_{2}\to hl$ (with lepton number $L_{1}$ and
branching ratio $r$) and $N_{2}\to \tilde{h}\bar{l}$ (with lepton
number $L_{2}$ and branching ratio $1-r$), and $C$ is a factor of
order unity quantifying the conversion of lepton number into baryon
number by sphalerons. The exact prediction for $M_{2}$ depends on the
amount of CP violation in the neutrino sector (and hence the
difference between the branching ratio $r$ and the branching ratio
$\bar{r}$ for the corresponding process involving anti-particles). We
leave the more detailed calculation of the parameter constraints from
this leptogenesis scenario for future work. For the time being, we
just note the following: independent of the details of CP violation in
the neutrino sector, we can rewrite the constraint on $M_{2}$ as a
lower-bound, as shown in (\ref{M2_formula}). 

This bound once again leads to tension with MDSM$_{1}$ and MDSM$_{2}$
since, in those models, one can barely make $M_{2}$ bigger than
$100~{\rm GeV}$ before destabilizing the Coleman-Weinberg potential.
For example, if we assume $m_{\perp}<500~{\rm GeV}$ (corresponding to
$\lambda_{h}<2$), stability of the Coleman-Weinberg potential in
Models 1 and 2 requires $M_{2}<400~{\rm GeV}$.  But, once again,
MDSM$_{3}$ escapes a similar fate, since one has the freedom to make
$M_{2}$ very large, as long as the $Z'$ mass is also very large.

Let us recapitulate what we have learned thus far.  Although all three
version of the MDSM may be made compatible with current collider
physics, MDSM$_{1}$ and MDSM$_{2}$ are {\it not} good candidates to
{\it also} drive a period of inflation in the early universe.  By
contrast, MDSM$_{3}$ {\it can} also drive a period of inflation in the
early universe; and then, the decay of the inflaton can directly
generate a non-thermal cold dark matter particle and leptogenesis, in
a rather elegant and economical fashion.

%%%%%%%%%%%%%%%%%%%%%%%%%%%%%%%%%%%%%%%%%%%%%%%%%%%%%%%%
\section{Dark matter and leptogenesis: other possibilities}
\label{DarkLeptoViaOther}

In the preceding section, we have described a scenario in which dark
matter and leptogenesis are both generated directly, and
non-thermally, via inflaton decay; and we have seen that, although
this scenario meshes nicely with MDSM$_{3}$, it is in conflict/tension
with MDSM$_{1}$ and MDSM$_{2}$.  In this section, we wish to briefly
draw the reader's attention to possible alternative routes to dark
matter and leptogenesis in these models.  These alternative routes are
important to mention, since they are not linked to a period of
primordial inflation driven by $\psi_{\parallel}$, and hence do not
lead to the same tension with MDSM$_{1}$ and MDSM$_{2}$.

Instead of producing heavy neutrinos via $\rho_{\parallel}$ decay
after inflation, we can produce them via oscillations from left-handed
neutrinos in the primordial thermal bath.  The heavy neutrinos
produced in this way can account for dark matter, leptogenesis, or
both.  In this picture, lightest of the three heavy neutrinos is a
warm dark matter candidate with a mass of a few keV.  For an analysis
of these scenarios, see \cite{Asaka:2005an, Asaka:2005pn}.

In Model 2, there is yet another possibility: Refs.\
\cite{Meissner:2008gj, Latosinski:2010qm} argue that the Majoran field
$a$ obtains a mass via quantum effects, and behaves just like an
axion.  Just like the usual axion, this particle can offer a solution
to the strong CP problem, and also act as a dark matter candidate
(see \cite{KolbTurner, Weinberg, Mukhanov}).

%%%%%%%%%%%%%%%%%%%%%%%%%%%%%%%%%%%%%%%%%%%%%%%%%%%%%%%%
\section{Compelling features of MDSM$_{3}$}
\label{MDSM3_is_nice}

In this section, we draw together a variety of nice features of
MDSM$_{3}$, to emphasize that it seems to be a particularly compelling
model.  In particular, the arguments in Subsections \ref{anomalies}
and Subsections \ref{raison_detre} have not been mentioned yet in this
paper; we know that they are not original to us, but we do not know
what the correct original reference is, and would appreciate it if
some reader could point us to it.

\subsection{Extra gauge symmetry, non-trivial anomaly cancellation}
\label{anomalies}

An interesting theoretical route by which we can arrive at the model
MDSM$_{3}$ is the following.  Suppose we look for an extension of the
standard model with the following properties.  We want it to have the
same fermion content as the standard model (including right-handed
neutrinos); and we want those fermions to be grouped into 3 identical
generations, as in the standard model.  We also want the theory to
include the usual Yukawa terms that couple the fermions to the Higgs
doublet $h$.  The fields will have the same $SU(3)\times SU(2)$
representations as in the standard model, but now we want to extend
the gauge group to $SU(3)\times SU(2)\times U(1)_{1}\times\ldots\times
U(1)_{n}$, so that, instead of containing a single $U(1)$ factor, it
now contains $n$ different $U(1)$ factors.  Let us start by allowing
the various fields to carry arbitrary charges under these various
$U(1)$ gauge groups, and see what the above conditions imply.  Thus,
we have the following fields, with the following charges under
$SU(3)\times SU(2)\times U(1)_{1}\times\ldots\times U(1)_{n}$:
\begin{equation}
  \label{anomalies_table}
  \begin{array}{c|c|c|l|l|l} 
    & SU(3)_{C} & SU(2)_{L} & U(1)_{1} & \ldots & U(1)_{n} \\
    \hline
    q_{L} & 3 & 2 & \;Q_{1,q} & \ldots & \;Q_{n,q} \\
    \hline
    u_{R} & 3 & 1 & \;Q_{1,u} & \ldots & \;Q_{n,u} \\
    \hline
    d_{R} & 3 & 1 & \;Q_{1,d} & \ldots & \;Q_{n,d} \\
    \hline
    l_{L} & 1 & 2 & \;Q_{1,l} & \ldots & \;Q_{n,l} \\
    \hline
    \nu_{R} & 1 & 1 & \;Q_{1,\nu} & \ldots & \;Q_{n,\nu} \\
    \hline
    e_{R} & 1 & 1 & \;Q_{1,e} & \ldots & \;Q_{n,e} \\
    \hline
    h & 1 & 2 & \;Q_{1,h} & \ldots & \;Q_{n,h}
  \end{array}
\end{equation}
The standard model contains Yukawa terms like
\begin{subequations}
  \begin{eqnarray}
    &\bar{q}_{L}\tilde{h}\,u_{R}+h.c.& \\
    &\bar{q}_{L}h\,d_{R}+h.c. & \\
    &\bar{l}_{L}\tilde{h}\,\nu_{R}+h.c.& \\
    &\bar{l}_{L}h\,e_{R}+h.c.& 
  \end{eqnarray}
\end{subequations}
For the extended model to retain these terms, they must remain gauge
invariant, which implies the following conditions:
\begin{subequations}
  \label{yukawa_gauge_inv2}
  \begin{eqnarray}
    0&=&-Q_{i,q}+Q_{i,u}-Q_{i,h} \\
    0&=&-Q_{i,q}+Q_{i,d}+Q_{i,h} \\
    0&=&-Q_{i,l}+Q_{i,\nu}-Q_{i,h} \\
    0&=&-Q_{i,l}+Q_{i,e}+Q_{i,h}
  \end{eqnarray}
\end{subequations}
for all $i\in[1,n]$.  A number of additional constraints come from 
requirement that the chiral and gravitational anomalies vanish; the
constraints involving the $U(1)'s$ are:
\begin{subequations}
  \begin{eqnarray}
    A[U(1)_{i}]&=0=&
    3[6Q_{i,q}\!-\!3Q_{i,u}\!-\!3Q_{i,d}\!+\!2Q_{i,l}\!-\!Q_{i,\nu}\!-\!Q_{i,e}]\\
    A[U(1)_{i}U(1)_{j}U(1)_{k}]&=0=&
    3[6Q_{i,q}Q_{j,q}Q_{k,q}\!-\!3Q_{i,u}Q_{j,u}Q_{k,u}\!-\!3Q_{i,d}Q_{j,d}Q_{k,d}
    \qquad\nonumber\\
    &&\;
    +2Q_{i,l}Q_{j,l}Q_{k,l}\!-\!Q_{i,\nu}Q_{j,\nu}Q_{k,\nu}\!-\!Q_{i,e}Q_{j,e}Q_{k,e}]\\
    A[SU(3)^{2}U(1)_{i}]&=0=&
    \frac{3}{2}[2Q_{i,q}-Q_{i,u}-Q_{i,d}]\\
    A[SU(2)^{2}U(1)_{i}]&=0=&
    \frac{3}{2}[3Q_{i,q}+Q_{i,l}]
  \end{eqnarray}
\end{subequations}
for all $\{i,j,k\}\in[1,n]$.  It may be shown by induction that the
general solution is
\begin{equation}
  Q_{i,A}=\alpha_{i}Y_{A}+\beta_{i}(B-L)_{A}.
\end{equation}
In other words, each $U(1)$ charge can be a different linear
combination of ordinary hypercharge $Y$ and baryon-minus-lepton number
$(B-L)$.  What is the interpretation of this result?  

First of all, it tells us that only two of the $U(1)$'s are linearly
independent; the rest are redundant.  Thus we will henceforth assume
just two $U(1)$'s.  Then we can, without loss of generality, redefine
our fields so that the first $U(1)$ factor is purely hypercharge $Y$,
and the second $U(1)$ factor is purely $(B-L)$; this is the convention
we have followed in this paper.  

Secondly, the system is overconstrained: it is suprising to find a
solution at all, let alone one with so many free parameters.  In the
case of interest, $n=2$, we have 18 equations for 14 unknowns, and
find a solution with 2 free parameters.  We may be inclined to
interpret this apparent coincidence as theoretical evidence that the
additional $U(1)_{B-L}$ gauge symmetry of MDSM$_{3}$ is on the right
track.  [The vanishing of the $U(1)_{B-L}$ gauge anomaly in the
standard model (with right handed neutrinos) is, of course, well known,
and may also be intepreted/understood in terms of the fact that the
standard model may be embedded in a grand unified theory based on
$SO(10)$.]

\subsection{A {\it raison d'etre} for right-handed neutrinos}
\label{raison_detre}

The ordinary standard model makes sense, with or without the
right-handed neutrino.  In particular, since the right-handed neutrino
is a singlet under the standard model gauge group, it does not
contribute to anomaly cancellation: in the standard model, the
anomalies cancel whether or not the right-handed neutrino exists.  The
right-handed neutrino has recently been tacked on to the standard
model in order to account for the observed neutrino oscillations --
but not for any independent theoretical reason.

It is important to emphasize that the situation is very different in
MDSM$_{3}$.  In this case, the right-handed neutrino is charged under
$U(1)_{B-L}$, and thus plays an essential role in the anomaly
cancellation arguement in the previous subsection.  Stated another
way: if we tried to add an extra $U(1)$ gauge symmetry to the {\it
  minimal} standard model (with no right handed neutrinos), in the
manner described in the previous section, we would find that we
couldn't do it -- there would be no solutions to the corresponding
constraints.  We would be led to add a fermion to each generation,
with exactly the properties of the right-handed neutrino, in order to
make the anomalies cancel.

In this sense, MDSM$_{3}$ has more explanatory power than MDSM$_{1}$,
MDSM$_{2}$, or the standard model: it gives the right-handed neutrino
a genuine theoretical {\it raison d'etre}.

\subsection{One scalar, many different roles}

Relative to the standard model, MDSM$_{3}$ contains two new fields:
the complex scalar field $\varphi$ and the $(B-L)$ gauge field
$C_{\mu}$.  In this subsection, we would like to stress the elegantly
economical way in which $\varphi$ simultaneously accomplishes several
important phenomenological tasks in this model.

On the one hand, its properties -- {\it i.e.}\ its $SU(3)\times
SU(2)\times U(1)_{Y}\times U(1)_{B-L}$ charges shown in
(\ref{MDSM3_table}) -- were determined by requirement that it should
be able to provide a Majorana-like Yukawa coupling to right-handed
neutrinos, in order to give a see-saw mechanism for neutrino mass.  On
the other hand, these same properties are precisely what is needed in
order for $\varphi$ to perform another crucial task: spontaneously
break $(B-L)$ symmetry, and give mass to the $C_{\mu}$ boson via the
Higgs mechanism, without leaving any additional unwanted Goldstone
bosons.  

At the same time, we have seen that $\varphi$ expands the scalar
sector in a way that leads to successful electroweak symmetry breaking
via the Coleman-Weinberg mechanism, an inflaton candidate, a dark
matter candidate, and a possible explanation for the cosmological
matter/anti-matter asymmetry.

Finally, this model contains 4 different quantities that all must be
rather large, for very different phenomenological/cosmological
reasons: (i) the mass of the $Z_{\mu}'$ boson, (ii) the masses of the
heavy neutrinos; (iii) the mass of the $\rho_{\parallel}$ boson, and
(iv) the inflaton VEV.  And yet, in MDSM$_{3}$, all 4 large
dimensionful quantities have a common origin in the large VEV of the
field $\varphi$.

\subsection{Cosmology}

Finally, as we argued, MDSM$_{3}$ seems to be a viable extension of
the standard model of particle physics and, at the same time, seems
able to perform a wider variety of important cosmological tasks than
MDSM$_{1}$, MDSM$_{2}$, or the standard model.

Taken together, these reasons point to MDSM$_{3}$ as a particularly
interesting and compelling extension of the standard model.

%%%%%%%%%%%%%%%%%%%%%%%%%%%%%%%%%%%%%%%%%%%%%%%%%%%%%%%%
\section{Discussion}
\label{discuss}

Finally, let us mention some of the most important ways that we hope
to improve and extend this analysis in future work. (i) First, we plan
to perform a more complete analysis of leptogenesis, to flesh out the
treatment given above. (ii) Second, in this paper we have used the
Gildener-Weinberg formalism; but it would be better (especially in our
analysis of inflation) to perform a more complete analysis,
incorporating all renormalization-group effects, and using the full
renormalization-group-improved effective potential. (iii) Third, we
plan to examine the RG flow to find regions in parameter space that
are ``asymptotically safe'' (or, at least, free of instabilities or
Landau poles up to the Planck scale). (iv) Finally, in this paper we
have taken the MDSM to be {\it minimally} coupled to gravity; this
minimal coupling introduces a dimensionful constant (Newton's
gravitational constant $G$) which is arguably at odds with the
original spirit (and classical conformal invariance) of the original
non-gravitational MDSM. Thus, we are exploring the possibility that
there are other couplings to gravity that are more aligned with the
spirit of the MDSM, but do not spoil the most desirable features of
the model that follow from the minimally-coupled analysis performed in
this paper.

\acknowledgments

We are grateful to William Bardeen, Stephen Hawking, Philip Schuster,
Natalia Toro, Michael Trott, Neil Turok, James Wells, Mark Wise and
Itay Yavin for valuable conversations.  Research at Perimeter
Institute is supported by the Government of Canada through Industry
Canada and by the Province of Ontario through the Ministry of Research
\& Innovation. LB also acknowledges support from an CIFAR Junior
Fellowship.

%Appendix
\appendix

\end{document}